\documentclass{elsarticle}
\usepackage{graphicx}
\usepackage{amsmath}
\usepackage{amssymb}

\begin{document}

\title{Phantom dark energy and cosmological solutions without the Big Bang singularity}

\author[bau]{A.N.~Baushev}
\ead{baushev@gmail.com}
\address[bau]{Bogoliubov Laboratory of Theoretical Physics, Joint Institute for Nuclear Research\\
141980, Joliot-Curie str. 6, Dubna, Moscow Region, Russia}

\begin{abstract}
The hypothesis is rapidly gaining popularity that the dark energy pervading our universe is
extra-repulsive ($-p>\rho$). The density of such a substance (usually called phantom energy) grows
with the cosmological expansion and may become infinite in a finite time producing a Big Rip. In
this paper we analyze the late stages of the universe evolution and demonstrate that the presence
of the phantom energy in the universe is not enough in itself to produce the Big Rip. This
singularity occurrence requires the fulfillment of some additional, rather strong conditions. A
more probable outcome of the cosmological evolution is the decay of the phantom field into 'normal'
matter.

The second, more intriguing consequence of the presence of the phantom field is the possibility to
introduce a cosmological scenario that does not contain a Big Bang. In the framework of this model
the universe eternally expands, while its density and other physical parameters oscillate over a
wide range, never reaching the Plank values. Thus, the universe evolution has no singularities at
all.
\end{abstract}


\maketitle

\section{Introduction}

Experimental data \cite{wmap, riess, perlmutter} indicate convincingly an accelerating cosmological
expansion. The substance responsible for this effect is usually called dark energy. If we speculate
that the universe is filled with a perfect fluid with equation of state $p=w \rho$, the
accelerating expansion appears if $w<-1/3$. There are a lot of  models of the dark energy: it
can be explained by a non-zero cosmological constant (probably, this is the most natural
hypothesis, in this case $w=-1$) or by presence of a global cosmic scalar field slowly moving
down to the potential minimum (then $-1/3>w>-1$). Until recently, the instance of $w <
-1$ did not attract considerable interest. A substance with such an equation of state (it is
usually called phantom energy) generally possesses physically unacceptable properties, for example,
the sound speed in it can be super-luminal.

Nevertheless the case of $w<-1$ has currently come to the attention. First of all, it does not
contradict to the experimental data \cite{caldwell1, caldwell2}. Furthermore, it has turned out
that small perturbations are never super-luminal at least in some substances with $w<-1$ (see
\cite{gibbons} as an example). So, a phantom energy theory can be causal.

Physical properties of a dark energy with $w < -1$ differ markedly from the case when $w
\ge -1$. For instance, it violates the energy domination condition $\rho\ge |p|$. Since this is one
of the principal assumptions of the classical black hole non-diminishing theorem, a black hole mass
diminishes by accretion of the phantom energy \cite{babichev}. But the difference most important
for the cosmology becomes apparent when we consider expansion of the universe containing dark
energy. If $w \ge -1$ the dark energy density is not increasing (or even decreasing) as the
universe expands. All bound systems (such as the solar one) remain bound forever; the dark energy
presence manifests itself only on cosmological scales. If we allow $w<-1$,  the dark energy
density grows and becomes infinite in a finite time. Increasing gravitational repulsion produced by
the phantom energy will first destroy Galaxies and then any bound systems including elementary
particles \cite{caldwell1, caldwell2}. This phenomenon is usually called Big Rip. However, as we
will see below, realization of such a catastrophic scenario requires the fulfilment of several
supplementary conditions. Even if the phantom energy prevails in the universe, the Big Rip does not
necessarily occurs.

There is another less evident consequence of $w < -1$: if dark energy is really the phantom
one, the universe evolution needs not to contain a Big Bang.

\section{The phantom energy-dominated universe evolution}
The metric of a homogeneous isotropic universe can be represented as (hereafter we use the system
of units where $c=\hbar=1$, $8 \pi G = M_p^{-2}$)
\begin{eqnarray}
ds^2 = dt^2 - a^{2}(t)\,[\,dx^2 + dy^2 + dz^2\,] = \nonumber \\
= a^{2}(\eta)\,[\,d\eta^2 - dx^2 - dy^2 - dz^2\,] \label{a1}
\end{eqnarray}
 As we have already mentioned,
presence of the phantom energy in the universe is not enough in itself to produce the Big Rip. This
phenomenon appears, for instance, if the coefficient averaged over the universe contents $w =
{\it const} < -1$. However, it is beyond reason to believe that $w$ remains constant during
the cosmological evolution. In order to illustrate this let us at first consider a very simple
situation when a universe contains only uniformly distributed phantom energy with an ordinary
Lagrangian
\begin{equation}
\label{a2} \mathcal L = -\dfrac{\partial_\xi\!\phi\, \partial^\xi\!\phi}{2} - V(\phi)
\end{equation}
For such a field
\begin{equation}
\label{a3} \rho = -{\dot \phi^2}/2 + V(\phi), \quad p = -{\dot \phi^2}/2 - V(\phi)
\end{equation}
Hereafter a dot over a symbol denotes derivative with respect to time $t$. We obtain:
\begin{equation}
w \equiv \frac{p}{\rho} = \dfrac{\left(-{\dot \phi^2}/2 - V(\phi)\right)}{\left(-{\dot
\phi^2}/2 + V(\phi)\right)} \label{a4}
\end{equation}
One can see that $w$ is not a constant: it depends on the amplitude of the field and on the
potential energy density $V(\phi)$. This question has been discussed in close detail in
\cite{samitoporensky}. To summarize briefly the results: if the potential is not very steep (grows
slower than $V(\phi)\propto \phi^4$) then $w$ tends to $-1$, and the density becomes infinite
only when $t\to\infty$. So, no Big Rip appears in this case, though the universe reaches the Planck
density in a finite time. For steeper potentials a big rip singularity appears even if $w\to
-1$. Even the parameter $w$ can tend to $-\infty$ for a very steep $V(\phi)$. Finally,  if
$V(\phi)$ has a maximum, $w\to -1$ and $\rho\to {\it const}$ \cite{caroll2003,singh2003}. It
is important to emphasize that a very steep potential is necessary to provide a constant $w <
-1$: for any polynomial potential, for instance, $w$ tends to $-1$.

As an illustration let us consider the simplest case of $V(\phi)= m^2 \phi^2 /2$. The Einstein
equation can be written as
\begin{equation}
H^2 = \dfrac{\left(-{\dot \phi}^2 + m^2 \phi^2\right)}{6 M^2_p} \label{a5}
\end{equation}
Then the equation of motion $\ddot\phi + 3H\dot\phi - V'(\phi)=0$ takes the form
\begin{equation}
\ddot\phi + 3H\dot\phi - m^2 \phi=0 \label{a6}
\end{equation}
Initially the phantom field $\phi$ rapidly grows as well as the Hubble constant $H$ in accordance
with (\ref{a5}). In due course, the friction term $3H\dot\phi$ becomes very large, and the phantom
field remains almost constant. For this regime we have $\ddot \phi \ll 3H\dot\phi$, ${\dot \phi}^2
\ll m^2 \phi^2$, and (\ref{a5},\ref{a6}) can be rewritten as
\begin{equation}
\dot\phi \simeq m M_p \sqrt{\dfrac23}, \quad H \simeq \dfrac{m}{M_p} \dfrac{\phi}{\sqrt6}\label{a7}
\end{equation}
It is easy to see that no Big Rip occurs and $w\to -1$. Moreover, equations (\ref{a7}) are
closely analogous to those describing the chaotic inflation in the early universe \cite{lindeinfl}.
In both the instances the scalar field and its energy density changes very slowly, and the universe
evolves in a quasi-De-Sitter regime. The only difference is that in the case of chaotic inflation
the field slowly decreases, while in the considering case it slowly increases (usually it is called
'slow-climb').

In case of another potential $V(\phi)$ the University evolution dynamics can be quite different, in
particular, the slow-climb may be lacking. Rapid universe expansion, however, is a common feature
of all phantom-containing models. Indeed, the density of the phantom matter (contrary to the
'normal' one) grows with time, and finally it begins to dominate. Then even in the softest case of
$w = -1$ (that is not exactly a phantom energy) the scalar factor $a$ grows exponentially; if
$V(\phi)$ is steep, the expansion can be much faster. So, if the universe contains phantom energy,
it eventually comes to an inflation-like stage.

The second effect that is able to prevent the Big Rip is a possible phantom energy interaction
generating 'normal' particles. The normal substance gives positive contribution to the pressure
that decreases the $w$ value averaged over the universe contents. Eventually the Big Rip may
be prevented.

It is generally believed that the phantom energy does not interact with 'normal matter', but it
cannot be absolutely true, if for no other reason than the gravitational interaction. Let us
consider a toy illustration for a start. As noted above, in the phantom scenario increasing
gravitational repulsion produced by the phantom energy will finally destroy even elementary
particles. We consider a hadron (more precisely, a quark-antiquark pair $q\bar q$) ripped by the
gravitational forces \footnote{This process becomes possible when the energy of the gravitational
repulsion between the quarks in the hadron becomes comparable with the typical strong interaction
energy scale 1~{GeV}. The phantom energy density at this moment depends, in particular, upon the phantom energy model.
The gravitational repulsion force between two quarks can be estimated as
$$
f_g\sim (1+3w)\dfrac{4\pi G}{3} \rho m R
$$
where $m$ is the constituent quark mass, $R$ is the typical hadron radius.
Using the condition $f_g R\sim 1$~{GeV} and taking $m=300$~{MeV}, $R=10^{-13}$~{cm} and $w\simeq -1$, we obtain
for the density of the phantom energy $\rho\sim 10^{54}$~{g/$\text{cm}^3$}}.
When the distance between the quark and antiquark increases, the color field
lines of force between them are pressed together into a string-like region.  The gravitational
repulsion stretches the color lines of force until the increasing potential energy becomes
sufficient to create another $q\bar q$ pair. It divides the string on two strongly-coupled $q\bar
q$ pairs, but they are also ripped up by the gravitational repulsion. So, this leads to intensive
hadron production in the moments just before the Big Rip.

Another mechanism of 'normal' matter generation in the phantom-dominated universe is particle
production in the cosmological gravitational field. In the modern universe this effect is
completely negligible, but if the universe is phantom-dominated, its density (and, consequently,
the gravitational field) grows with time making the particle production very intensive. In
\cite{particleproduction} the process was considered under assumption that the universe is filled
with a perfect fluid with $w = {\it const} < -1$. The influence of the 'normal' matter on the
universe expansion was neglected. It was shown that if the conformal time $\eta$ is chosen so that
$\eta<0$, and the density becomes infinitive when $\eta \to -0$, the generated 'normal' matter
density can be written as
\begin{equation}
\rho_{norm}=C\eta^\beta, \quad\text{where} \quad\beta= \dfrac{4}{1+3w}\label{a8}
\end{equation}
In order to examine the problem in more detail we consider a universe filled with a perfect fluid
with $w=-4/3$. Initially we neglect the particle production. Then the cosmological equations
can be easily solved. It is convenient to choose the following initial conditions: $\rho_{de0}$,
$a_0=\sqrt{\dfrac{3}{8 \pi G \rho_{de0}}}$, $\eta_0=-\dfrac23$ (we denote the phantom energy density
by $\rho_{de}$ and its initial value by $\rho_{de0}$). The value of $\eta_0$ is taken so that the
density becomes infinitive when $\eta \to -0$. We obtain
\begin{equation}
\rho_{de}\propto \eta^{-\frac23}\label{a9}
 \end{equation}
 Now the particle production can be added with the help of (\ref{a8}). We will consider the
generated substance to be relativistic, i.e. with the equation of state $p_{norm}=\rho_{norm}/3$.
Then the system of cosmological equations can be written as
\begin{eqnarray}
\left(\dfrac{1}{a^2}\dfrac{da}{d\eta}\right)^2&=&\dfrac{1}{3 M^2_p} (\rho_{de}+C\eta^{-\frac43})\label{a10}\\
\dfrac{d(\rho_{de}+C\eta^{-\frac43})}{d\eta}&=&\dfrac{1}{a}\dfrac{da}{d\eta} (\rho_{de} - 4 C\eta^{-\frac43})\label{a11}\\
dt&=&a d\eta\label{a12}
\end{eqnarray}
The Hubble constant is given by the equation
\begin{equation}
H\equiv\dfrac{1}{a}\dfrac{da}{dt}=\dfrac{1}{a^2}\dfrac{da}{d\eta}\label{b1}
\end{equation}
 There is no accurate estimation of the $C$ value, but we can avoid this
difficulty choosing properly the initial density $\rho_{de0}$. Indeed, comparing (\ref{a8}) and
(\ref{a9}), one can see that the density of the 'normal' matter grows faster, and eventually it
becomes comparable with the phantom one. This is the moment of our interest. Therefore, it is
reasonable to take $\rho_{de0}= 100 \cdot C\eta^{-\frac43}_0$, i.e. to choose the initial moment so
that the initial phantom density is only hundred times higher then the 'normal' one. Since the
particle production is a weak effect, this value of $\rho_{de0}$ must be quite high.

\begin{figure}
 \resizebox{\hsize}{!}{\includegraphics[angle=90]{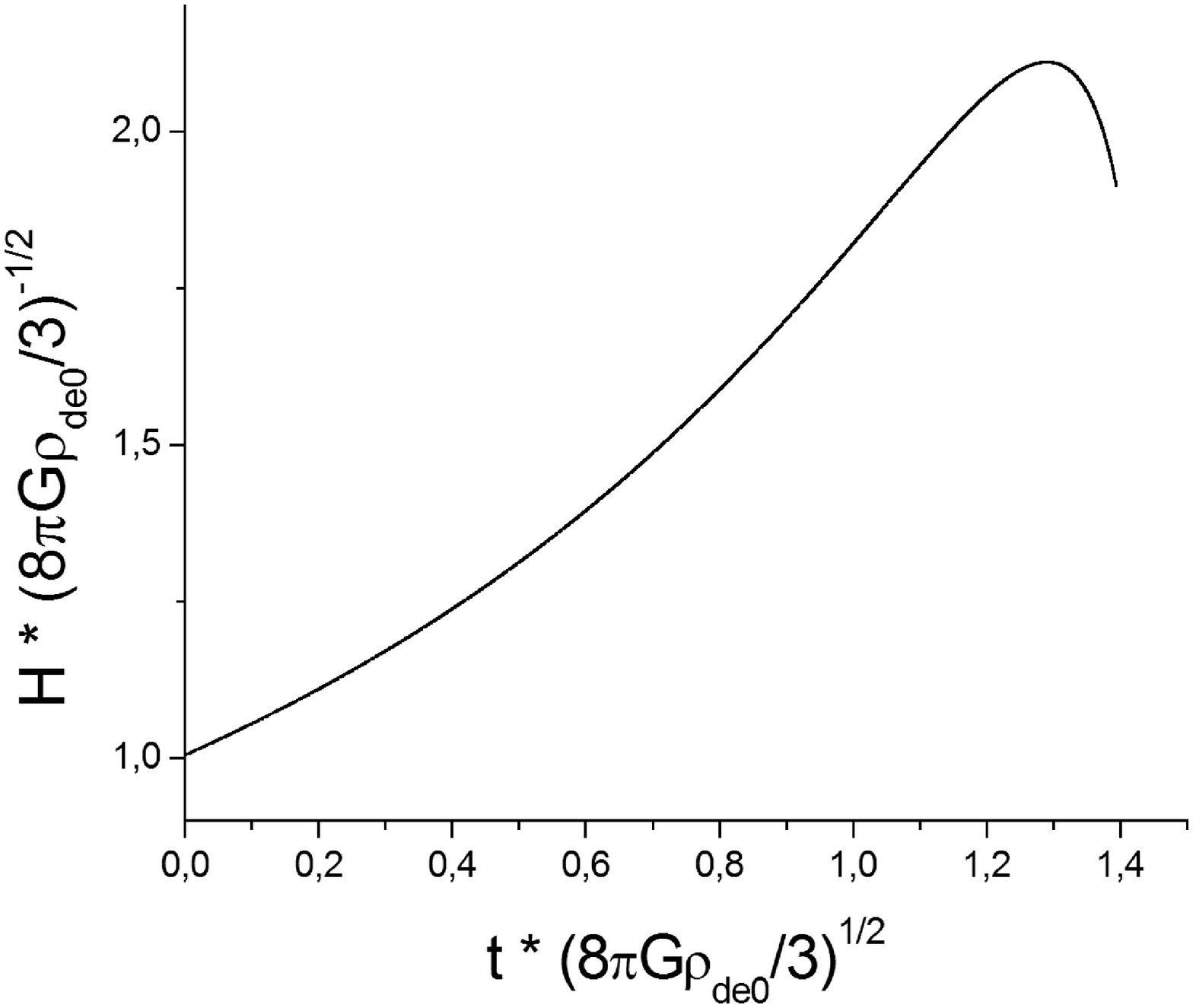}}
 \caption{The time dependence of the Hubble constant $H$. We use $H\cdot(8\pi G \rho_{de0}/3)^{-1/2}$ instead of $H$
 and $t\cdot(8\pi G \rho_{de0}/3)^{1/2}$ instead of $t$.}
 \label{fig1}
\end{figure}

\begin{figure}
 \resizebox{\hsize}{!}{\includegraphics[angle=270]{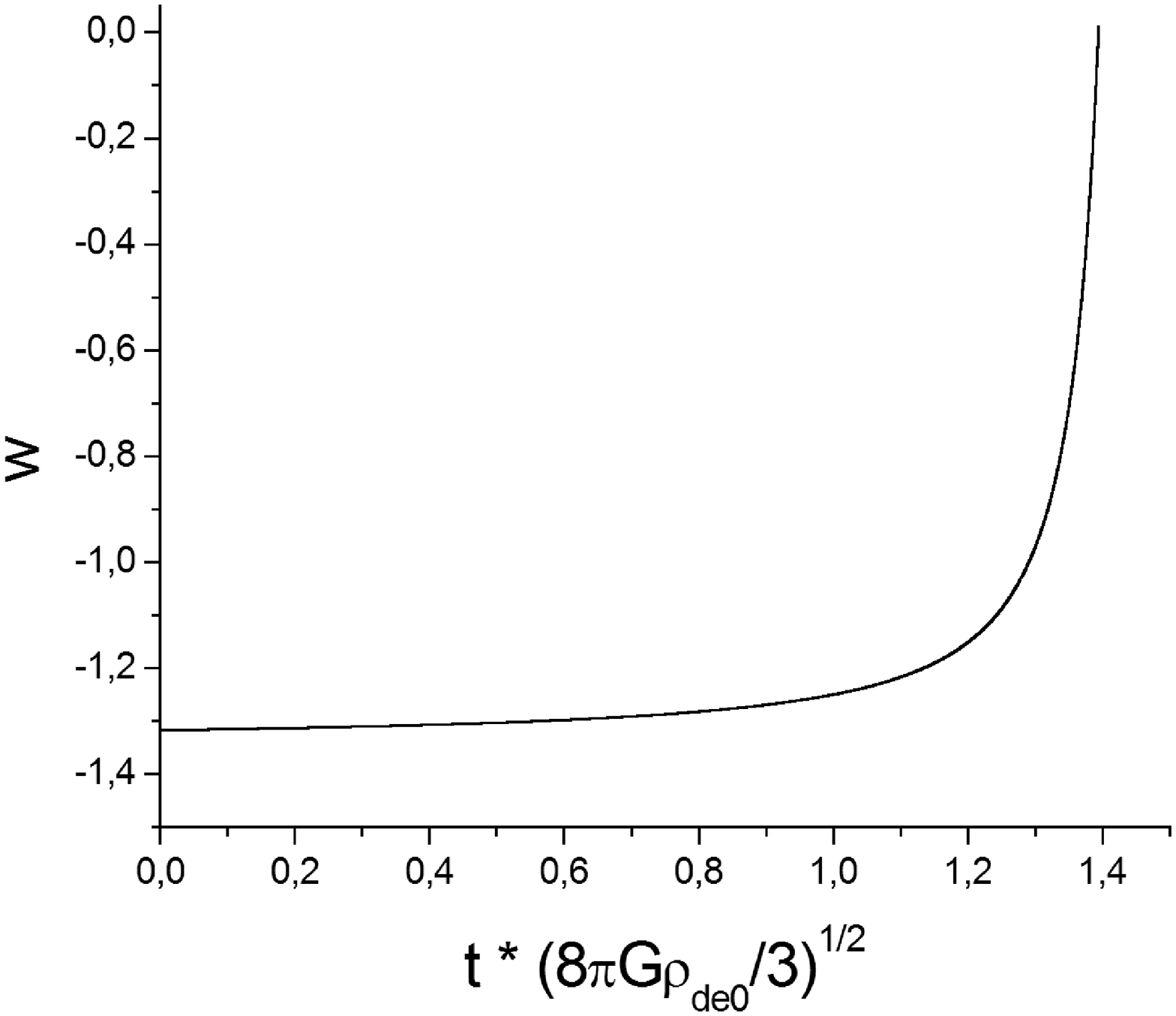}}
 \caption{The time dependence of $w$. We use $t\cdot(8\pi G \rho_{de0}/3)^{1/2}$ instead of $t$.}
 \label{fig2}
\end{figure}

Results of the (\ref{a10}-\ref{a12}) system integration with the aforecited boundary conditions are
represented on Fig.~\ref{fig1}-\ref{fig2}. It is convenient to use $H\cdot(8\pi G
\rho_{de0}/3)^{-1/2}$ instead of $H$ and $t\cdot(8\pi G \rho_{de0}/3)^{1/2}$ instead of $t$ in order to
avoid the dependence of the graphs from the uncertain parameter $\rho_{de0}$. It is seen from
Fig.~\ref{fig1} that the Hubble constant grows progressively slower, reaches its maximum $H_{max}$,
and begins to decrease. On the contrary,  $w$ constantly increases from the initial value that
is close to $-4/3$. Finally it becomes positive. Of course, both the effects result from the high
growth fraction of the 'normal' matter in the universe that makes a positive contribution to the
pressure. One can see from Fig.~\ref{fig1} that $H_{max}\simeq 2.11\cdot(8\pi G \rho_{de0}/3)^{1/2}$.
As we have already mentioned, $\rho_{de0}$ is a very big quantity, so, $H_{max}$ must be also quite
large.

Extension of the solution to $w=0$ makes no sense: the equation (\ref{a8}) is obtained under
the assumption that the dynamics of the universe expansion is determined only by the phantom
energy. It becomes inapplicable when the fraction of 'normal' matter gets noticeable. It seems
reasonable to contend, however, that in the course of the phantom-dominated universe evolution the
quantity of 'normal' matter becomes comparable with the phantom one, and the value of $w$
grows hindering from the Big Rip. To this we can add that we have considered the case of
$w={\it const}$, which claims an extremely steep phantom field potential $V(\phi)$, as we
could see above. For a more realistic case of the polynomial potential $w$ tends to zero even
in the absence of 'normal matter'. This makes the occurrence of the Big Rip highly unlikely.

\section{Discussion}
Of course, the suggested model is ingenuous. Intensive phantom field may interact in some other way
decaying into 'normal' matter. Whatever the mechanism of the transformation may be, the universe
after it has the following properties:
\begin{enumerate}
 \item It has just passed through the stage of very rapid (at least, exponential) expansion.
 \item It is flat and homogeneous.
 \item It is 'normal' matter-dominated.
 \item The Hubble constant is very large.
\end{enumerate}
It is precisely these physical conditions that existed in our universe $\sim 13.7$ milliard years
ago, just after the inflation. If we assume that the inflation was caused by the phantom field (as
we could see, the equations of the phantom cosmology (\ref{a7}) are able to be closely analogous to
those of the chaotic inflation \cite{lindeinfl}), this makes possible to construct a model of
universe evolution that does not contain a Big Bang. The universe leaves inflation being
matter-dominated. As it expands the density of matter decreases, while the phantom density grows:
eventually the universe passes into the phantom-dominated stage. As this takes place, the total
density and the Hubble constant stop diminishing and begin to increase. Gradually the expansion
becomes very fast, leading to an inflation-lake stage. Finally the phantom field decays into
'normal' matter, and the cycle repeats. Thus, the universe eternally expands, while its density and
other physical parameters oscillate over a wide range, never reaching the Plank values. The
universe evolution has no singularities like a Big Bang or a Big Rip.

 The foregoing model of self-reproducing cyclic universe evolution is, in a certain sense, a
return to the Steady State theory \cite{hoyle1,hoyle2,hoyle3,gibbons}. Indeed, in both the cases
the universe expands forever with no Big Bang, and the matter is generated by the decay of some
field (in our case this is the phantom field). The difference is that in the classical Steady State
universe this process goes on constantly providing constant universe density, while in the
considering case the universe evolves cyclically returning to its original state, but the density
varies within wide limits. Presence of the evolution allows to avoid most difficulties inherent to
the classical Steady State model. Thus, the considering model may be called the Steady State {\it
on the average} universe.

The presence of the inflation-like stage in the considering scenario solves most of the
problems appearing in an inflationless cosmological theory \cite{lindeinfl}. The size of the
universe during this stage increases so much that an observer in the after-inflation epoch
can see only a small part of it. Consequently, the universe looks very flat and homogeneous.
Moreover, as we could see (\ref{a7}), it is possible to choose the model of the phantom energy so that even
the dynamics of the inflation-like stage is similar to the standard chaotic inflation scenario \cite{lindeinfl}.

Is the proposed model valid for the real universe? The answer to this question is currently not
known with certainty, primarily because of the absence of direct experimental proofs of the phantom
matter existence. A non-zero cosmological constant is quite enough to explain the experimental data.
On the other hand, observational data show that allowed values of the parameter
$w$ of the modern universe (at $2\sigma$) lie between $-2.25 $ and $\sim -0.5$
\cite{caldwell2}. So, the existence of the phantom energy is not experimentally forbidden.

Theoretical status of the phantom energy is also vague. There are a lot of Lagrangians known, which
describe fields with the necessary properties. It is still unclear, however, whether it is possible
to introduce such a Lagrangian, completely free from paradoxical physical effects like
super-luminal sound speed, catastrophic quantum instability of the vacuum etc.

Nevertheless, if the phantom energy does exist, the considering model is an interesting instance of
cosmological solutions containing no singular points like a Big Bang or a Big Rip. It is worthy of
note that, though in our case the universe density never reaches infinity (or, more precisely,
Planck density), all the physical parameters vary over a very wide range and amount up to extremal
values. So, the universe passes through the stages that can be called physically critical.

\section{Acknowledgements} This work was supported by a grant from the RFBR (Russian Foundation for Basic
Research, Grant 08-02-00856).

\end{document}